\documentstyle[prb,aps,twocolumn,epsf]{revtex}

\begin{document}
  
\draft
\date{\today}  
\title{ Optical Absorption of CuO$_3$ antiferromagnetic chains at finite 
temperatures}
\author{Eduardo Gagliano, Fabian Lema, Silvia Bacci, Juan Jos\'e
Vicente}
\address{Comisi\'on Nacional de Energ{\'\i}a At\'omica,
Centro At\'omico Bariloche and Instituto Balseiro,\\
8400 S.C. de Bariloche (RN), Argentina.}
\author{J. Lorenzana\cite{lore}}
\address{ Istituto Nazionale di Fisica della Materia - Dipartimento 
di Fisica,\\ Universit\`a di Roma ``La Sapienza'', Piazzale 
A. Moro 2, I-00185 Roma}

\date{\today}

\maketitle
\begin{abstract}
We use a high-statistic quantum Monte Carlo  and Maximum Entropy  
regularization method to compute the dynamical energy correlation 
function (DECF) of the one-dimensional (1D) $S=1/2$ antiferromagnetic 
Heisenberg model at finite 
temperatures. We also present a finite  temperature analytical ansatz for the 
DECF which is in very good agreement with the numerical 
data in all the considered temperature range. From these results, and
from a finite temperature generalisation of the  
mechanism proposed by Lorenzana and Sawatsky [Phys.\ Rev.\ Lett. 
{\bf 74}, 1867 (1995)], we compute the line shape for the optical 
absorption spectra of multimagnon  excitations assisted by phonons 
for quasi 1D compounds. The line shape has two contributions
analogous to the Stokes and anti-Stokes process of Raman scattering. 
Our low temperature data is in good agreement with optical 
absorption experiments of CuO$_3$ chains in Sr$_2$CuO$_3$.  Our finite 
temperature results provide a non trivial prediction on the dynamics of 
the Heisenberg model at finite temperatures that is easy to verify
 experimentally. 
\end{abstract}

\narrowtext

\section{  INTRODUCTION  }
Infrared optical absorption measurements\cite{suz96} on charge-transfer  
insulators such  as Sr$_2$CuO$_3$ and the isostructural compound  
Ca$_2$CuO$_3$ have shown  that these experiments provide important  
insights in our understanding of the  spin dynamics of the underlying  
CuO$_3$ chains. The observed bands are analogous to the bands found in
cuprates with CuO planes\cite{per93,per94,per98,cho98} at mid infrared 
energies.  

Although direct absorption of magnetic excitations is not 
possible in the usual optical transitions induced by dipole interaction, 
they are allowed by the assistance of phonons\cite{lor95,lor95b,lor97,lor99} 
whose effect is to effectively lower the symmetry of the lattice giving 
rise to a finite dipole moment\cite{koide}.
This effective dipole moment for processes involving one phonon and spin  
excitations has been obtained perturbatively starting from the three-band
Hubbard model in the presence of slowly varying electric and phonon
fields (see Ref.~\onlinecite{lor95,lor95b} hereafter LS) or using exact  
diagonalization of small clusters\cite{gar98}.

An effective low-energy description of the spin degrees of freedom of the
CuO$_3$ chain is given by the one-dimensional (1D) $S=1/2$ 
antiferromagnetic Heisenberg Hamiltonian (AFH).
\begin{equation}
H_{{\rm mag}}=  J \sum_{i} {\bbox{S}_{i}\bbox{S}_{i+1}}.
  \label{h}
\end{equation}
Although the energy spectrum of this many-body quantum problem has been 
solved a long time ago\cite{bet31}, the computation of its dynamical 
properties remains a challenge. So far most of the numerical results on
 dynamical 
properties come from exact diagonalization of small systems, and more 
recently
from quantum Monte Carlo (QMC) simulations combined with the Maximum 
Entropy (MaxEnt) 
analytic continuation procedure of noisy data. This combination has
 been used 
to calculate the dynamical spin correlation  
function of the 1D\cite{dei,sta97} and 2D\cite{Makiv,Greven} Heisenberg model
 and correctly describes the  
inelastic neutron scattering of CuCl$_2$.2N(C$_5$H$_5$)\cite{dei} and  
La$_2$CuO$_4$\cite{Makiv,Greven}.  
  
The estimated intrachain exchange energy of Sr$_2$CuO$_3$ is  
$J\sim 0.26 eV$\cite{suz96,lor97,ami95}.
For an electric field parallel to the CuO$_3$ chain the absorption band
at low temperatures ($T= 32K$) is wide, starts at the relevant phonon 
frequency  
($\omega_0= 0.07$eV), extends up to $\omega=\omega_0+\pi J \sim 0.9$eV
(the maximum energy of two spinons dispersion) and has a singularity at  
$\omega= \omega_0+\pi J/2 \sim 0.48$eV (the maximum of the des 
Cloizeaux-Pearson dispersion). For an electric field perpendicular to 
the chain axis, the spectrum is weak and structureless. This kind of 
response is typical of 1D systems.  

Experimental features\cite{suz96} of the optical absorption  
spectrum $\alpha(\omega)$ of these materials may be 
analysed applying LS idea of phonon-assisted multimagnon optical 
absorption with a $T= 0$ analytical ansatz for the dynamical energy 
correlation  function (DECF)\cite{lor97}. 
Numerical studies show that at $T= 0$ the dynamical  
structure factor and the DECF of the
AFH model have a significant structure in the $\omega-q$ plane between  
$\omega_1(q)= \pi J|\sin(q)|/2$ (the des Cloizeaux-Pearson dispersion 
relation) and $\omega_2(q)= \pi J|\sin(q)|$, the maximum of the 
two-spinon continuum. At zero temperature the LS mechanism gives structure 
at the energy of the relevant phonon plus the energy of the magnetic 
excitations and the computed spectrum is in excellent agreement 
with the experimental one\cite{lor97} at low temperatures.

Here we present  numerical results for the {\em finite temperature}  
DECF.  For systems up to $N_{s}= 32$, we  
perform a high-statistic quantum Monte Carlo simulation and obtain the  
spectrum by the MaxEnt analytic continuation regularization  
technique of noisy data. The Monte Carlo technique is a non-perturbative  
approach and so useful to check the validity of the different 
approximations involved on the analytical results; the considered system 
sizes are large enough to safely extrapolate to the thermodynamic limit.
The technique is presented on Sec.~\ref{qm}.  

One of our goals is to study finite temperature effects on the optical 
absorption  spectra. Although the temperature dependence of the spectra
has been studied experimentally in 2D compounds\cite{per98,cho98} 
 to the best of our knowledge no 1D studies have been reported yet.
 In Sec.~\ref{ft} we discuss LS theory at finite temperatures. 

 We also check  
the validity of the $T= 0$ ansatz\cite{lor97} for the DECF and, 
based on bosonization results\cite{cro79},
we generalise this ansatz to finite temperatures (Sec. \ref{an}). 
The analytical ansatz is in very good 
agreement with the numerical data in all the considered temperature 
range. Finally, our results for the optical absorption spectra (Sec. \ref{ir})
are in excellent agreement with available low temperature experimental
(Ref. \onlinecite{lor97}) data
and provide a prediction on the finite temperature dynamics of a 1D Heisenberg
system that can be tested experimentally.

\section{PHONON-ASSISTED IR ABSORPTION OF MAGNONS AT FINITE TEMPERATURES}
\label{ft}
\subsection{Model}

In this section we generalise  the $T= 0$ theory of LS to finite  
temperatures.  We consider an effective Hamiltonian describing the coupling 
of light with one-phonon-multimagnon excitations: $H= H_{0}+H_{1}$, where  
$H_{1}= -PE$, $P$ ($E$) is the magnitude of the dipole moment operator 
(electric field) in the  chain direction. The unperturbed 
Hamiltonian is:  
\begin{equation}
H_{0}= H_{{\rm mag}}+H_{{\rm ph}}.  \label{h0}
\end{equation}
here the magnetic part is the Heisenberg model Eq.~(\ref{h}) and for the  
phonon Hamiltonian we take a single branch: the Cu-O stretching mode in 
the  chain direction\cite{lor95}   
\begin{equation}
H_{{\rm ph}}= \sum_{q }\omega _{q }(a_{q }^{\dag }a_{q }+ \frac{1}{2})
\end{equation}
where $\omega _{q }$ is the phonon frequency. The dipole moment is  
\begin{equation}
P= \sum_{q }\lambda _{q }A_{q }B_{-q .}  \label{pab}
\end{equation}
where $A_{q }= (a_{q }+a_{-q }^{\dag })$, $B_{q }$ is
the Fourier transform of $B_{i}\equiv\bbox{S}_{i}\bbox{S}_{i+1}$.  
The strength of the coupling of light with these excitations is given by:
\begin{equation}
\lambda _{k }= 4q_{{\rm A}}\sqrt{\frac{\hbar}{2M\omega _{k }}}\sin^{2}(\frac{k }{2})
\end{equation}
here $M$ is the oxygen mass and $q_{{\rm A}}$ a material dependent effective 
charge which is in the range  $0.1 \sim 0.4 e$ \cite{lor95,lor95b,lor97}.

In the definition of the dipole moment operator we have neglected terms like 
conventional phonon absorption ($P= \lambda^{\prime }A_{0}$) which are outside 
the scope of this work.

\subsection{\protect\bigskip Optical absorption}
  
From general considerations the optical absorption spectrum is given by
\begin{eqnarray}\label{alfa}
\alpha(\omega) = {{2\omega}\over c} {\rm Im}(\sqrt{ \epsilon(\omega) })
\end{eqnarray}
where $\epsilon= \epsilon_{1}(\omega)+i\epsilon_{2}(\omega)$ is 
the complex-dielectric function and $c$ is the speed of light. Assuming 
weak absorption, $\epsilon_{1}>>\epsilon_{2}$, Eq.~(\ref{alfa}) can be 
rewritten as,
\begin{eqnarray}\label{adw}
\alpha(\omega) =  { {4\pi}\over {c\sqrt{\epsilon_1}} }\sigma(\omega)
\end{eqnarray}
where $\sigma (\omega)$ is the frequency dependent conductivity which can
be written in terms of the Fourier transform of the real time 
dipole-dipole retarded Green function. Using
the Lehman representation and the fluctuation-dissipation theorem one 
can write for the real part,
\begin{equation}
\sigma = \frac{\pi \omega (1-e^{-\beta \omega })}{ZV}\sum_{N,M}e^{-\beta
\varepsilon _{N}}|\langle N|P|M\rangle |^{2}\delta (\omega +\varepsilon
_{N}-\varepsilon _{M})
\end{equation}
here $H_0 |M\rangle =  \varepsilon_{M} |M\rangle$, $Z$ is the partition  
function, $\beta$ the inverse temperature and $V$ the volume.
In Eq.~(\ref{h0}) weak magnon-phonon interactions were assumed, i.e in 
the absence of electric fields any term coupling magnons with phonons was 
neglected. Under this assumption the eigenstates of the system are products 
of phonon states times magnetic states,
\begin{equation}
|M\rangle = |m\rangle |\mu \rangle
\end{equation}
where $|\mu \rangle$ ($|m\rangle$) labels an eigenstate of $H_{\rm mag}$ 
($H_{\rm ph}$) with energy $E_{\mu}$ ($\omega_m$).
 (From now on we use $\mu,\nu$ to label magnetic eigenstates and 
$m,n$ for phonon eigenstates.)

The optical conductivity can be written as:
\begin{eqnarray}
\sigma =\frac{\pi \omega (1-e^{-\beta \omega})}{VZ_{{\rm mag}}Z_{{\rm ph}
}}\sum_{q mn\mu \nu }e^{-\beta (\omega_n+E_{\nu})} \lambda_q^2 \times\\
| \langle n|A_q|m\rangle |^2 | \langle \mu|B_q|\nu\rangle |^2
\delta (\omega+\omega_n- \omega_m+E_{\nu}-E_{\mu}) \nonumber
\end{eqnarray}
where $Z_{\rm ph}= \sum_{n}e^{-\beta \omega_n}$ and $Z_{\rm mag}= 
\sum_{\nu}e^{-\beta E_{\nu}}$. Using the fact that $A_q$ creates or
destroys a phonon of momentum $q$, we can evaluate the phonon part and we 
get for the optical conductivity,
\begin{eqnarray}
\sigma&&= \frac{\pi\omega(1-e^{-\beta \omega}) }{V}\times\nonumber\\
 &&\sum_q \lambda_q^2
[(1+n_q) J_q(\omega-\omega_q) + n_q J_q(\omega+\omega_q)]   \label{sdw}
\end{eqnarray} 
$n_{q}$ is the Bose occupation number for $q$-momentum phonons.
All the magnetic information is stored in the function,
\begin{equation}
 J_{q}(\omega )=\frac{1}{Z_{{\rm mag}}}\sum_{\mu \nu }e^{-\beta E_{\nu
}}\left| \langle \mu |B_{q }|\nu \rangle \right| ^{2}\delta (\omega
+E_{\nu }-E_{\mu })
\end{equation}

 The first 
term in the brackets in Eq. (\ref{sdw}) corresponds to process in
which the incoming photon creates a magnetic excitation and a phonon; the
second term corresponds to the destruction of a phonon from the bath with
creation of a magnetic excitation. The first term does not vanish at zero
temperature and it is the one considered by LS. It gives structure at the 
energy  
of the phonons plus the energy of the magnetic excitations. The last term 
is new, vanish at zero temperature and gives structure at the energy of 
the magnetic excitation minus the energy of the phonon. Notice the 
similarity with the Stokes and anti-Stokes processes in Raman scattering.

$J_{q}$ can be evaluated from the retarded Green function using the 
fluctuation-dissipation theorem:
\begin{equation}
J_{q}(\omega )=\frac{-
\mathop{\rm Im}G_{q}(\omega )}{\pi(1-e^{-\beta \omega })}.
\end{equation}
where $G_q(\omega )$ is the Fourier transform of the real time retarded Green 
function 
\begin{equation}
  \label{gqdw}
G_q(t)=-i\theta(t)<[B(q,t),B(-q,0)]>.
\end{equation}
 We also define
 \begin{equation}
   \label{lambda}
\Lambda_{q}(\omega)\equiv (1-e^{-\beta \omega}) J_q(\omega)=-\frac1{\pi}
\mathop{\rm Im}G_{q}(\omega )     
 \end{equation}
 which is an odd function of $\omega$ and is evaluated in the next section
using MaxEnt and in Sec.~\ref{an} using an analytical ansatz.

\section{  QUANTUM MONTE CARLO AND MAXIMUM-ENTROPY ANALYTIC 
CONTINUATION  }
\label{qm}

In order to compute the magnetic response we perform a World Line 
quantum Monte Carlo simulation.\cite{hirsch}
The algorithm is based on a path integral approach. To evaluate
the average of physical observables,  
$$\langle O \rangle=
\frac{{\rm tr}(O e^{-\beta H_{\rm mag}})}{Z_{\rm mag}},$$
the method divides the $0< \tau < \beta$ interval into L parts, each one 
of width $\Delta \tau = \beta/L$ and breaks the 1D Hamiltonian 
$H_{\rm mag}$ in two  
commuting parts $H_1$ and $ H_2$, each one corresponding to a different  
sublattice, even and odd respectively. Introducing at each one of the 2L  
intermediate times a complete set of $S_z$ eigenstates and using the  
Suzuki-Trotter formula\cite{Suzu}  
$$e^{-\Delta\tau H_{\rm mag}} =  e^{-\Delta\tau H_1} e^{-\Delta\tau H_2}  
[1+O(\Delta\tau^2)],$$ one can evaluate the resulting matrix elements and 
insure,  
making $\Delta \tau \rightarrow 0$, that the error introduced by the  
Suzuki-Trotter approximation is smaller than the statistical error. The 
updating procedure has two different kinds of moves: local changes 
conserve the total spin projection  
$S_z$ and global moves allow magnetisation fluctuations.\cite{marcu}

We implemented a multispin code. This allows an efficient integer 
algebra manipulation of one-bit spins packed into 32-bits words and 
achieves a high speed. For different values of $\beta= 1,5,10$ (in units 
of J) we fix $\Delta \tau=  0.125$; given this value of $\Delta \tau$ 
the acceptance rates of local and global moves are always less than 
$10\%$, this is fixed by the algorithm and can not be changed.  
For $\beta= 10$ we performed $\sim 100$ independent runs, $10^6$ 
iterations each, measuring every 1000. It was necessary to evaluate the 
spin autocorrelation function in order to insure that measurements were 
not correlated.   
  
Quantum Monte Carlo simulations allow us to compute imaginary-time $\tau$
correlation functions. Besides the DECF results
presented in this work, as a check of our code, we have also computed  
static properties, the spin dynamic structure factor and for small 
systems, up to $N_s= 16$ sites, tested against exact diagonalization\cite{gag90}  
obtaining the optical absorption spectra from the spectral representation 
of the DECF. All
the results shown here correspond to a system of size $N_s = 32$.  
For low temperatures, $\beta= 10$ and $N_s = 32$ we monitored  
$\langle E \rangle^2  \sim \langle E^2\rangle  
= \frac{1}{N_s} \sum_{l= 0,...,N_{s}-1} {<B_i B_{i+l}>}$: in all  
cases the differences were less than $5 \times 10^{-4}$.

The imaginary time energy correlation function is periodic in $\tau$
with period $\beta$ and is given by
\begin{eqnarray}\label{gqtau}
G_q(\tau) =  \langle B_q(\tau) B_{-q}(0)\rangle
\end{eqnarray}
where $B_q(\tau)= e^{\tau H} B_q e^{-\tau H}$ and $0< \tau< \beta$.

For the analytic continuation problem of calculating the spectral 
density $\Lambda_q(\omega)$ associated with the imaginary-time
Green's function $G_q(\tau)$ one has to invert the integral equation,
\begin{eqnarray}
G_q(\tau) = \int_{-\infty}^{+\infty} d\omega K(\tau,\omega)
\Lambda_q(\omega)  
\end{eqnarray}
where $K$ is the Kernel for bosons
\begin{eqnarray}
K(\tau,\omega) = {{e^{-\tau \omega}}\over {{1-e^{-\beta \omega}}}}
\end{eqnarray}
It is well known that Fredholm integral equations of the first kind are  
classical examples of ill-posed problems. After discretization of the
frequency interval one ends up with a discrete ill-posed problem.  
Because QMC simulations provide a noisy and incomplete set of data  
${\bar G}_q(\tau_n)$, $\tau_n = n\Delta \tau$ the direct inversion 
is ill-conditioned and the $\Lambda_q(\omega)$ spectrum cannot be 
uniquely determined. Here the bar over $G_q(\tau_n)$ indicates a Monte 
Carlo average as explained below.

 The MaxEnt approach is a regularization method 
for this kind of ill-posed problems which provides 
{\it an unique solution} compatible with Bayesian statistic and prior 
knowledge.\cite{max-general} The best image compatible with the data is  
obtained maximising $$Q = \lambda S - {1\over 2} \chi^{2}$$ by solving
$\nabla Q= 0$. Here $\lambda$ is a regularization parameter and $\chi^{2}$ 
the misfit statistic which not only includes the statistical errors of  
the data but also the correlations between different imaginary-time data  
points.
The misfit statistic is given by
\begin{eqnarray}
\chi^{2} = \sum_{i,j} [G_q(\tau_i)-{{\bar G}_q(\tau_i)}]
C^{-1}_{ij}[G_q(\tau_j)-{{\bar G}_q(\tau_j)}]  
\end{eqnarray}
 where $C$ is the covariance matrix that takes into account 
the imaginary-time correlation of the data. $S$ is the Shannon-Jaynes  
entropy of the spectrum defined with respect to the initial guess model
$m(\omega)$:
\FL
\begin{equation}
S =- \int_{-\infty}^{+\infty} d\omega [\Lambda_q(\omega)-m(\omega)-
\Lambda_q(\omega) ln(\Lambda_q(\omega)/m(\omega))]
\end{equation}
$m(\omega)$ is the best guess in absence of prior knowledge on  
the spectrum. We took a flat model.

In practice, functional maximisation is reduced to search in a  
multidimensional parameter space by a Simpson evaluation of the 
integrals. We performed a variable $\omega$-discretization with 
100-200 frequency points.\cite{brian84}

\begin{figure}[tbp]
\epsfverbosetrue
\epsfxsize=8cm
$$
%\epsfbox[20 350 580 750]{gaustest.eps}
\epsfbox[20 10 580 750]{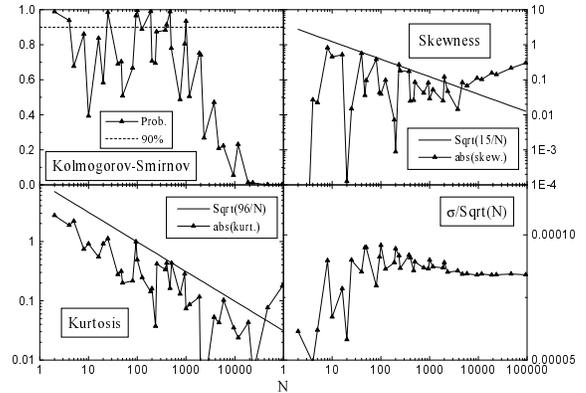}
%\epsfbox[100 590 450 790]{gaustest.eps}
%\epsfbox{gaustest.eps}
$$
\caption{ The 4 different gaussianity tests that we  
performed on our data. The independent variable 
is $N$ the number of bins.  For Kolmogorov-Smirnov we plot 
the probability of the data being Gaussian.  
We can see that  up to $N= 1000$ bins there is a high probability of 
the data being Gaussian. For Skewness and  Kurtosis
crossing over the  full line indicates departure from Gaussianity. 
From the standard deviation ($\sigma$) test we can see
that under $N= 100$ bins the data does not behave as  Gaussian.
}
\label{gaustest}
\end{figure}

We binned the data $G_q$ until we obtained a Gaussian distribution that
satisfies several criteria, Kolmogorov-Smirnov, skewness, Kurtosis and  
standard deviation.\cite{test-gauss} Fig.~\ref{gaustest} presents an example: 
starting from $\sim 100000$ QMC-data the four gaussianity tests are 
satisfied grouping the data on $100-1000$ bins, each one an average of 
$1000-100$ QMC-data respectively. We consider $N= 1000$ bins. After this 
binning the imaginary time Monte Carlo averages ${\bar G}_q$ have 
relative errors less than $10^{-3}$.

A zero temperature sum rule\cite{lor97} was used to check the low  temperature
data and was found to be satisfied within a  $2\%$ error.

\section{Ansatz for the dynamic energy correlation function at
finite temperatures}
 \label{an}     
We present here a finite temperatures generalisation of the $T=0$ 
ansatz (Ref.~\onlinecite{lor97}), 
which agrees very well with the numerical results. 

The $T = 0$ ansatz   can be written in terms
of $G^{XY'}_q(\omega)$, the retarded Green function of the $XY$ model,
\begin{equation}
-\frac1{\pi}\mathop{\rm Im}G^{XY'}_q(\omega)= 
\frac{\theta [\omega -\omega_1(q)]\theta [\omega_2(q)-\omega ]
\sqrt{\omega_2^2(q)-\omega^2}}{\omega_2^2(q)}
\label{imgxy}
\end{equation}
here the prime indicates that the energy scale has been changed 
according to $J_{XY}\rightarrow J\pi /2 $.  
With this notation the zero temperature ansatz reads
\begin{equation}
\mathop{\rm Im} G_q(\omega )= \mathop{\rm Im} G^{XY'}_q(\omega)
[ A \sqrt{J f_{\pi}(\omega) }+B J f_q(\omega) ]
\label{igdwt}
\end{equation}
where A and B are constants of values A= 2.4, B= 0.6 and we defined  
\begin{equation}\label{fdq}
f_q(\omega)= \frac1{\sqrt{\omega^2-\omega_1^2(q)}}.
\end{equation}
The term $1/{\sqrt{\omega^2-\omega_1^2(q)}} $ in 
Eqs.~(\ref{igdwt}),(\ref{fdq})
is chosen to reproduce the behaviour of bosonization at low energies. 
For details see Ref. ~\onlinecite{lor97}.  

To generalise this ansatz to finite temperatures we simply replace
$G^{XY'}_q(\omega)$ and $f_q(\omega)$ by their finite temperature 
versions.
At finite temperature it is trivial to compute $G^{XY'}_q(\omega)$ 
by mapping the $XY$ model to a free Fermion model\cite{sol79},
\begin{equation}
\mathop{\rm Im}G^{XY'}_q(\omega)= -\frac{\sinh(\frac{\beta\omega}{2})
\mathop{\rm Re} \sqrt{\omega_2^2(q)-\omega^2}} {\omega_2^2(q)
\left[ \cosh(\frac{\beta\omega}{2}) +  
\cosh(\frac{\beta \sqrt{\omega_2^2(q)-\omega^2}}{2\tan(\frac{q}{2})})
\right] }  \label{imgxyt}
\end{equation}
On the other hand the low energy behaviour predicted by bosonization
becomes\cite{cro79}:
\begin{equation}
f_q(\omega)= -2 \mathop{\rm Im}  
(\beta I_1[\beta\frac{\omega-\omega_2(q)}{2\pi}]  
       I_1[\beta\frac{\omega+\omega_2(q)}{2\pi}])
\end{equation}
where we have defined
\begin{equation}\label{i1dy}
I_1(y)= \frac1{\sqrt{8\pi}}\frac{\Gamma(\frac14+\frac12 i y)}
                               {\Gamma(\frac34+\frac12 i y)}  
\end{equation}
Eqs.~(\ref{igdwt}),(\ref{imgxyt})-(\ref{i1dy}) define our finite 
temperature ansatz. The spectral density $\Lambda$ is computed from the
Green function through expression Eq.~(\ref{lambda}).  

\begin{figure}[tbp]
\epsfverbosetrue
\epsfysize=6cm
$$
\epsfbox[50 475 500 725]{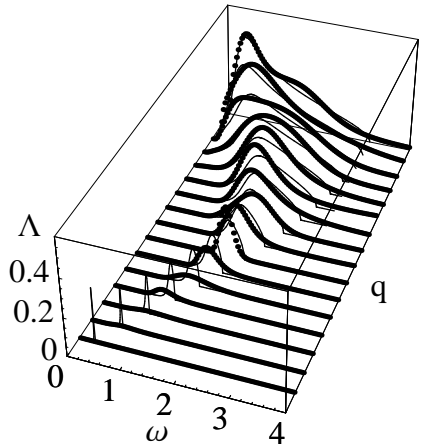}
$$
\caption{The spectral density
$\Lambda$ as a function of $q$ and $\omega$ for $\beta= 10$ ($J=1$). 
The dots are the MaxEnt data for $N_s= 32$ and the lines 
the finite temperature analytic ansatz.
}
\label{ldpwb10}
\end{figure}
\begin{figure}[tbp]
\epsfverbosetrue
\epsfxsize=10cm
$$
\epsfbox[100 590 450 790]{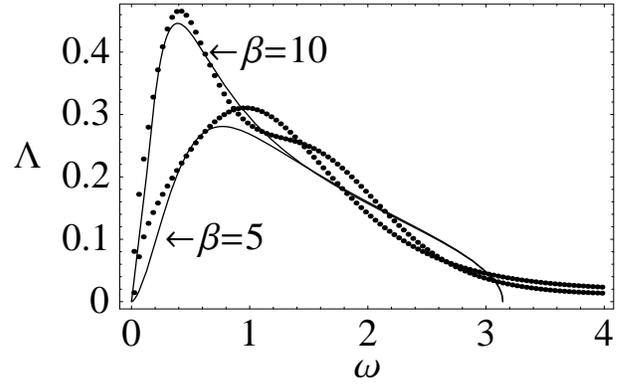}
$$
\caption{  $\Lambda$ as function of 
$\omega$ for $q= \pi$ and  $\beta=5,10$ ($J=1$).
 Dots and lines as in Fig.~\protect\ref{ldpwb10}.
}
\label{ldw}
\end{figure}

In Fig.~\ref{ldpwb10} we compare the finite temperature 
ansatz with the Maximum 
Entropy data for different momenta and temperatures. We see that both 
approaches give quite similar results. In particular the width of the 
peak close to $q= \pi$ agrees remarkably well (See Fig.~\ref{ldw}).
 For  $q \sim 0$ the ansatz gives a much narrower peak than the MaxEnt data, 
however this disagreement is of no importance for our purpose since 
this part of the spectrum gives a negligible contribution to the IR line 
shape.  

\section{Infrared absorption spectra}
\label{ir}
\begin{figure}[tbp]
\epsfverbosetrue
\epsfxsize=10cm
$$
\epsfbox[100 590 450 790]{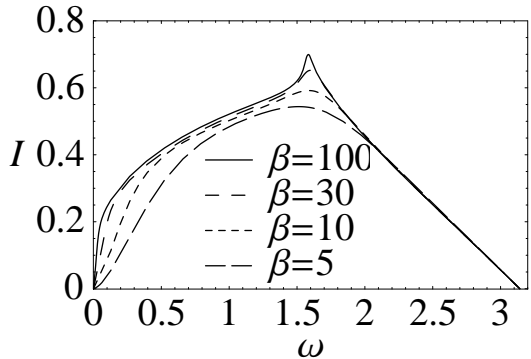}
$$\caption{ $I(\omega)$ for different values of $\beta$
in units such that $J=1$.}
\label{idw}
\end{figure}

The optical spectra for phonon-assisted multimagnon absorption is given 
by Eqs.~(\ref{adw}),(\ref{sdw}). For the materials under consideration 
the  momentum dependence of the phonons can be neglected and we can take 
Einstein phonons: $\omega_q\equiv\omega_0$, $n_q\equiv n_0$. In this case 
it is  useful to define the following function,
\begin{equation}
I(\omega)= -\frac{8}{N_s\pi} \sum_{q} \sin^4(q/2) 
\mathop{\rm Im} G_{q}(\omega ) ,
\end{equation}
In Fig.~\ref{idw}  we show this function for different values of the inverse 
temperature. From this figure the absorption 
coefficient for other materials can be computed without the cumbersome 
integration of the ansatz. The absorption coefficient for a specific 
material is,
\begin{eqnarray}
\alpha =\alpha_0 \omega (1-e^{-\beta \omega }) 
[ \frac{(1+n_0)I(\omega-\omega_0)} {1-e^{-\beta(\omega-\omega_0)}}  
+ \frac{n_0 I(\omega+\omega_0)}{1-e^{-\beta(\omega+\omega_0)}}]  
\label{sdwe}
\end{eqnarray} 
where  
$$
\alpha_0= \frac{4 N_s \pi^2 q_{\rm A}^2}{c\sqrt{\epsilon_1}MV\omega_0}.
$$

\begin{figure}[tbp]
\epsfverbosetrue
\epsfxsize=10cm
$$
\epsfbox[100 590 450 790]{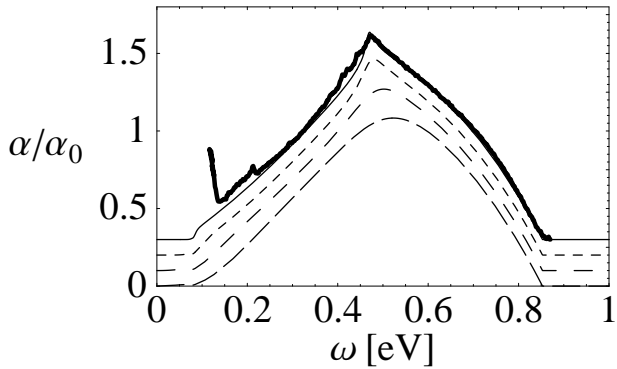}
$$
\caption{ We show $\alpha$ as a function of $\omega$
for Sr$_2$CuO$_3$ ($J=0.246$eV, $\omega_0=0.08$eV) 
  for (from bottom to top) $\beta/J=5,10,30,100$,
 using  the finite temperature ansatz.
 The curves are shifted by 0.1 for
clarity. The thick line is the experimental data after
Ref.~\protect\onlinecite{suz96}}
\label{sdw.fig}
\end{figure}

In Fig.~\ref{sdw.fig} 
we show our results for the midinfrared absorption spectra
obtained from the analytical ansatz  
and compare with experimental measurements on Sr$_2$CuO$_3$ at $T=32K$.
As a check we can see that 
our low temperature results are in excellent agreement with experimental 
data.
   
\section{  SUMMARY AND DISCUSSION }

We have used quantum Monte Carlo and MaxEnt methods to calculate
the dynamical energy correlation function, $\Lambda_{q}(\omega)$, for  
S= 1/2 antiferromagnetic Heisenberg chains at finite temperatures.  
  
Based on bosonization at finite temperatures we also present a finite  
temperature ansatz for the  spectral density of the DECF
which is in very good agreement with numerical data in all the 
temperature range. Since these two approaches are independent the 
agreement gives us confidence on the accuracy of both results. 
In particular for  
$q= \pi$, the agreement of the width of the spectral function is a 
quite strong test for the numerical data of finite systems because in 
the ansatz that width is completely determined by bosonization (which is 
an asintotically exact approach).

From  the ansatz we obtain the optical absorption which, 
at low temperatures, is in excellent agreement with  
available CuO$_3$ measurements. This is expected since the finite
temperature ansatz converges to the  zero temperature ansatz as the 
temperature is lowered and for the latter good agreement has already 
been found\cite{lor97}.

Naively one can expect that finite temperature effects will affect  the 
line shape at energies smaller than the temperature. As Fig.~\ref{sdw.fig} 
show 
this expectation is wrong. Dramatic changes occur close to the 
singularity at energy $\omega_0+\pi J/2$. 
This can be understood by considering 
the mapping of the problem to a half-filed interacting-fermion 
band\cite{sol79}. The particle hole transitions that have the required 
energy to contribute to the singularity are excitations of 
fermions from just below the Fermi level to the top of the band and 
particle hole transitions from the bottom of the band to just above the 
Fermi level. Since both process involve the fermions at the Fermi level
 they get strongly affected by the smoothing of the Fermi distribution function 
around the Fermi energy. This explains the rounding of the singularity 
with temperature. Of course the low energy part of the spectrum also involves
excitations close to the Fermi level and gets affected as can be seen 
from Fig.~\ref{idw}.  

Our finite temperature line shape is a non trivial prediction on the dynamics
of the 1D Heisenberg model at finite temperatures. 
We hope that this result will stimulate further experimental work in 
this direction.

\section{  ACKNOWLEDGEMENTS  }

All of us are supported by the Consejo Nacional de Investigaciones  
Cient\'{\i }ficas y T\'{e}cnicas, Argentina.  
Partial support from Fundaci\'on Antorchas under grant 13016/1,
from Agencia Nacional de Promoci\'on Cient\'{\i }fica y
Tecnol\'ogica under grant  PMT-PICT0005 and PICT 03-00121-02153 
and CONICET grant 4952/96 and INFM PRA (HTSC) are  gratefully 
acknowledged.  J. L. thanks the IRS group (Rome) for  hospitality.

\end{document}